\documentclass[aps,prb,twocolumn]{revtex4-1}

\usepackage{graphicx}
\usepackage{multirow}

\usepackage{amsmath}
\usepackage{color}
\usepackage{soul}

\begin{document}

\title{Orbital--dependent backflow wave functions for real--space quantum Monte Carlo}

\author{Markus Holzmann}
\affiliation{Univ. Grenoble Alpes, CNRS, LPMMC, 3800 Grenoble, France}
\affiliation{Institut Laue Langevin, BP 156, F-38042 Grenoble Cedex 9, France}

\author{Saverio Moroni}
\affiliation{CNR-IOM DEMOCRITOS, Istituto Officina dei Materiali, and SISSA Scuola Internazionale Superiore di Studi Avanzati, Via Bonomea 265, I-34136 Trieste, Italy}

\begin{abstract}
We present and motivate an efficient way to include orbital dependent many--body correlations
in trial wave function of real--space Quantum Monte Carlo methods 
for use in electronic structure calculations. We apply our new orbital--dependent backflow
wave function to calculate ground state energies of the first row atoms
using variational and diffusion Monte Carlo methods.
The systematic overall gain of correlation energy with 
respect to single determinant Jastrow-Slater wave functions is competitive with 
the best single determinant trial wave functions currently available.
The computational cost per Monte Carlo step is comparable to that of simple backflow calculations.
\end{abstract}


\maketitle

\section{Introduction}
The fermion sign problem in general prevents electronic quantum Monte Carlo (QMC) calculations
from determining unbiased ground--state properties within a controlled precision
and only polynomial increasing computational 
cost
in the number of particles. Real--space QMC methods\cite{reviews} avoid
the sign problem through the fixed--node approximation,
solving the Schr\"odinger equation with Dirichlet boundary 
conditions on the nodes of a trial function $\Psi$.
While fixed--node  results are often accurate, the quest for reducing
the systematic error incurred has prompted generalizations of
the standard Jastrow--Slater (JS) wave function. Better
wave functions are obtained replacing the Slater determinant 
by a multideterminant  expansion \cite{filippi}, antisymmetrized 
geminal product (AGP)\cite{sorella} or Pfaffian (PF)\cite{schmidt_prb}. As an
alternative or in addition, backflow (BF) transformations\cite{slkc} 
can be applied to the particles' coordinates. All these variations 
include correlation effects in the nodal structure of $\Psi$, which 
in turn determines the accuracy of the fixed--node approximation.

In this paper we introduce a way of including electron correlations
in the antisymmetric factor of $\Psi$ improving
the nodal structure of strongly inhomogeneous systems.
Whereas in previous BF wave functions the $i$th particle's
coordinate ${\bf r}_i$ in the argument of the $n$th single--particle orbital
is substituted
by the BF--transformed coordinate ${\bf q}_i$ [e.g. given by Eq.~(\ref{eq_eta}) below],
\begin{equation}
\phi_n({\bf r}_i)\longrightarrow\phi_n[{\bf q}_i(X)],
\label{eq_stdbf}
\end{equation}
where $X$ specifies the configuration of the system (e.g. electronic
and nuclear coordinates), we instead replace each orbital by two or more orbitals
coupled via BF correlations in the amplitudes,
\begin{equation}
\phi_n({\bf r}_i)\longrightarrow\phi_n^{(1)}({\bf r}_i)
+[{\bf q}_i(X)-{\bf r}_i]\cdot\nabla\phi_n^{(2)}({\bf r}_i).
\label{eq_obf}
\end{equation}
Here $\phi_n^{(a)}$, $a=1, 2, \dots$, denote reoptimized orbitals of the same spatial symmetry as 
$\phi_n$, specific to the $n$th orbital.
Thus, 
the same BF transformation ${\bf q}_i(X)$ affects differently the
various orbitals describing the antisymmetric part of $\Psi$.
We call "orbital backflow" (OBF) 
this way of using the transformed coordinates.

The OBF functional form is motivated in Sec. \ref{sec_formalism} using
the local energy method\cite{kwon,mh_bf} for a single--determinant wave
function, and applied to the first row atoms in Sec. \ref{sec_results},
where it proves competitive with inhomogeneous 
backflow (IBF)\cite{brown,seth},
AGP\cite{sorella,sorella_unpublished} and PF\cite{schmidt_prl,schmidt_prb} 
wave functions. 

\section{Orbital Backflow Trial Wave Function}
\label{sec_formalism}
We briefly outline how normal
and orbital backflow may emerge naturally from approximating 
a generalized Feynman-Kac path integral formula. We are merely interested in possible functional
forms, suitable for numerical evaluation, so that most of the approximations
in this section are
driven more by the need of simplifcation than by mathematical rigour. 
Thus, anticipating the eventual optimization of the functional parameters
of any resulting trial wave function, we use variational freedom already
in intermediate simplification steps, 
to modify some of the detailed expressions 
into plausible functional forms suggested by physical intuition.
The notation
$\widetilde{f}(\cdot)$ will be used  to indicate changes 
of an explicit function $f(\cdot)$ due to parameter optimization. 

The ratio between the exact ground--state wave function $\Phi(R)$ and a
trial wave function $\Psi_0(R)$ not orthgonal to $\Phi$ is\cite{kalos,caffarel,mh_bf}
\begin{equation}
	\frac{\Phi(R)}{\Psi_0(R)} \propto \langle e^{-\int_0^\infty E_L(R(t))dt}\rangle,
\label{eq_gfk}
\end{equation}
where $R=({\bf r}_1,\ldots,{\bf r}_N)$ are the coordinates of the
$N$ particles, and the brackets denote the average over all random
walks $R(t)$ starting at $R$ generated by the importance--sampled 
Green's function. The local energy method\cite{kwon,mh_bf} uses an 
analytic approximation of Eq. (\ref{eq_gfk}) to give an explicit 
expression for an improved wave function $\Psi$ in terms of $\Psi_0$
and its local energy $E_L(R)=\langle R|H|\Psi_0\rangle/\langle R|\Psi_0\rangle$,
\begin{equation}
\frac{\Phi(R)}{\Psi_0(R)} \approx e^{-\langle\int_0^\tau E_L(R(t))dt\rangle}
\approx e^{-\tau{\widetilde E}_L(R)}\equiv\frac{\Psi(R)}{\Psi_0(R)}.
\label{eq_elocal}
\end{equation}
The approximations underlying  Eq. (\ref{eq_elocal}) are the truncation 
of the cumulant expansion at first order over a finite projection time $\tau$,
and the assumption
that the random walk average of time integrals of $E_L[R(t)]$ 
merely reproduces the same functional form of the local energy,
but with a 
smoother $R$  dependence in the relevant phase-space region. 
The resulting expression ${\widetilde E}_L(R)$ in the exponent
of the improved wave function is therefore given by a functional expression
similar to $E_L(R)$ containing modified/optimized
pseudopotentials and orbitals.

We take $\Psi_0$ as a simple wave function with a Jastrow
factor $e^{-U(R)}$ and a Hartree product of single--particle orbitals 
$\phi_n({\bf r}_i)$ (the antisymmetrization being applied afterwards,
on the improved wave function $\Psi$). The modified local energy 
${\widetilde E}_L(R)$ then contains terms proportional to 
$\nabla_i {\widetilde U}(R) \cdot\nabla_i\ln{\widetilde \phi}_n({\bf r}_i)$.
Specializing further to a two--body Jastrow factor $U(R)=\sum_{i<j}u(r_{ij})$, 
Eq. (\ref{eq_elocal}) suggests that
the one--particle orbitals in the Slater determinant of 
the improved wave function $\Psi$ are given by
\begin{equation}
{\widetilde\phi}_n({\bf r}_i)\exp\left[\sum_{j\ne i}\frac{{\widetilde u}'}{r_{ij}}({\bf r}_i-{\bf r}_j)\cdot\nabla_i\ln{\widetilde \phi}_n({\bf r}_i)\right].
\label{eq_phitilde}
\end{equation}

When $\ln\phi_n$ is linear in ${\bf r}_i$, e.g.
$-i{\bf k}_n\cdot{\bf r}_i$ for plane waves of wave vector ${\bf k}_n$
describing homogeneous systems, we recover the 
familiar case of Eq. (\ref{eq_stdbf}) 
with the usual backflow transformation
\begin{equation}
{\bf q}_i={\bf r}_i+\sum_{j\ne i}\eta(r_{ij})({\bf r}_i-{\bf r}_j),
\label{eq_eta}
\end{equation}
where $\eta={\widetilde u}'/r_{ij}$.

Whereas the cumulant expansion in the local energy method guarantees the extensivity
of the logarithm of $\Psi$ for extended systems, this approximation may poorly describe
modifications of strongly inhomogeneous, localized orbitals.
Local modifications of orbitals may better be captured by
keeping only the linear term of the exponential of Eq. (\ref{eq_phitilde}). 
By further choosing different modified orbitals $\phi_n^{(a)}$ for each $n$,
to improve the variational flexibility of our trial wave function,
the OBF form of Eq. (\ref{eq_obf}) is obtained.
In our case, 
${\bf q}_i$ remains a simple backflow coordinate with homogeneous 
two--body correlations of the form given by Eq. (\ref{eq_eta}).

Let us stress that 
the sequence of approximations made to simplify
Eq. (\ref{eq_gfk}) are rather crude and remain on a heuristic level.
However, the procedure is not aimed to directly obtain accurate expressions, but
to suggest flexible functional forms for the trial 
function suitable for approximating the exact ground state within polynomial computational cost. 
The quality of the resulting functional form is determined {\em a posteriori} for specific systems after optimization of 
the radial function $\eta$ and the modified orbitals $\phi_n^{(a)}$ involved.

\section{Case study of the first row atoms}
\label{sec_results}
To benchmark the accuracy of the OBF wave function we have calculated the energies 
of all--electron first row atoms with variational Monte Carlo (VMC) and
fixed--node diffusion Monte Carlo (DMC) for a trial wave function represented by the product
of a Jastrow factor and a single--determinant per spin component composed
from backflow improved orbitals according to the transformation 
(\ref{eq_obf}).

In particular, $s$ orbitals now obtain the following form
\begin{equation}
\phi_n^s({\bf r}_i,{\bf q}_i)=\chi_n^{(1)}(r_i)
+({\bf q}_i-{\bf r}_i)\cdot{\bf r}_i\chi_n^{(2)}(r_i),
\label{s_orbitals}
\end{equation}
where ${\bf q}_i$ is given by Eq. (\ref{eq_eta}) using different $\eta$ 
functions for like-- and unlike-- spin electrons expressed as locally 
piecewise--quintic Hermite interpolants (LPQHI)\cite{natoli}, and
$\chi_j^{(\alpha)}$ are radial functions expanded in a basis of Slater
type orbitals\cite{roetti}.
The $p$ orbitals read
\begin{eqnarray}
\phi_n^{p_\alpha}({\bf r}_i,{\bf q}_i)&=&r^\alpha_i\chi_n^{(1)}(r_i)
+(q^\alpha_i-r^\alpha_i)\chi_n^{(2)}(r_i)\\ \nonumber
&+&r^\alpha_i({\bf q}_i-{\bf r}_i)\cdot{\bf r}_i\chi_n^{(3)}(r_i)
\label{p_orbitals}
\end{eqnarray}
where $\alpha$ is the cartesian component required in the $n$th orbital.
Instead of using $[\partial\chi_n^{(2)}(r)/\partial r]/r$ as suggested by
Eq. (\ref{eq_obf}), we introduced a third independent radial function
$\chi_n^{(3)}(r)$ 
for increased variational freedom.

Implementation of OBF is rather straightforward by considering both
the particles' coordinates ${\bf r}_i$ and the
renormalized BF coordinates ${\bf q}_i$ as independent variables of the modified orbitals on the 
right--hand side of
(\ref{eq_obf}). Gradient and Laplacian of the trial wave functions are then
obtained by applying the chain rule in the same way as for
standard BF \cite{kwon}. Compared to a 
direct inclusion of orbital--dependent BF correlations
through different coordinate transformations for different orbitals,
the computational cost of our OBF wave function thus
maintains the overall $N^3$ scaling of standard BF, with a small additional
cost of less than a factor 2.
The increased number of independent terms in each orbital can be dealt with
by modern optimization techniques \cite{umrigar,rocca}.

The symmetric Jastrow factor of our case study on first row atoms contains an electron--electron term 
$\prod_{i<j}\exp[-u(r_{ij})]$ with different pseudopotentials $u$
for like and unlike spins, an electron--nucleus term
$\prod_i\exp[-w(r_i)]$, and 
electron--electron--nucleus correlations
\begin{equation}
\prod_{i\ne j}\exp\{-[\xi_0(r_i)\xi_0(r_j)
-\xi_1(r_i)\xi_1(r_j){\bf r}_i\cdot{\bf r}_j]\}.
\label{eq_3body}
\end{equation}
All radial functions $u$, $w$, $\xi_0$, and $\xi_1$
are expressed as LPQHI.
The variational parameters (58 for Li and Be, 67 for the other atoms)
are optimized by minimization of the variational energy\cite{rocca}.
The resulting VMC and DMC energies obtained are listed in Table \ref{tab_table}.
\begin{table}[h]
	\begin{tabular}{|r|l|l|}
\hline
  Z &    ~~~~~~VMC          &   ~~~~~~DMC\\
\hline
  3 &~~-7.4777221(63)&~~-7.478002(25)\\
  4 &~-14.661198(15) &~-14.664801(90)\\
  5 &~-24.642090(24) &~-24.64840(24)\\
  6 &~-37.83091(13)  &~-37.83796(20)\\
  7 &~-54.576653(69) &~-54.58366(18)\\
  8 &~-75.05034(14)  &~-75.059814(96)\\
  9 &~-99.71480(19)  &~-99.72617(14)\\
 10 &-128.91956(21)  &-128.93129(28)\\
\hline
\end{tabular}
\caption{Energies in Hartree a.u. of the first row atoms 
obtained with VMC and fixed--node DMC using the OBF wave function.}
\label{tab_table}
\end{table}
\begin{figure}[h]
\includegraphics[width=\columnwidth]{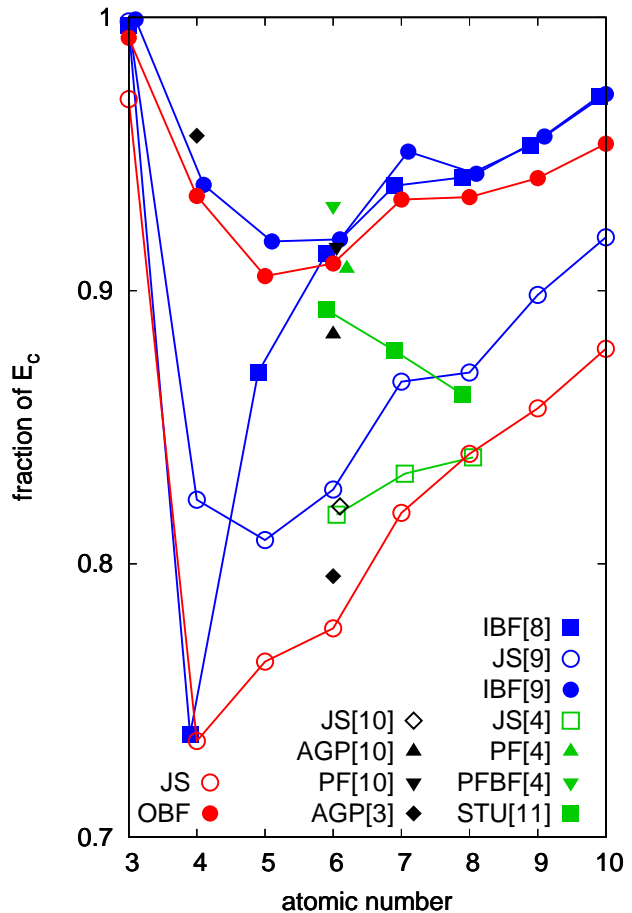}
\caption{
Fraction of correlation energy recovered in VMC for
the first row atoms of orbital backflow wave functions (OBF)
compared to previous results  using
different kinds of single--determinant wave function,
namely antisymmetrized
geminal product (AGP)\cite{sorella,sorella_unpublished},
Pfaffian (PF)\cite{schmidt_prb},
inhomogeneous backflow (IBF)\cite{brown,seth},
Pfaffian including inhomogeneous backflow (PFBF)\cite{schmidt_prb},
and a general Pfaffian form dubbed STU
\cite{schmidt_prl}. 
Empty symbols denote the bare Slater-Jastrow result from
the respective sources.
Small horizontal shifts of some data are added for clarity.
The Hartree--Fock and estimated exact energies are 
taken from Table I of Ref. \onlinecite{brown}. 
}
\label{fig_vmc}
\end{figure}

\begin{figure}[h]
\includegraphics[angle=270,width=\columnwidth]{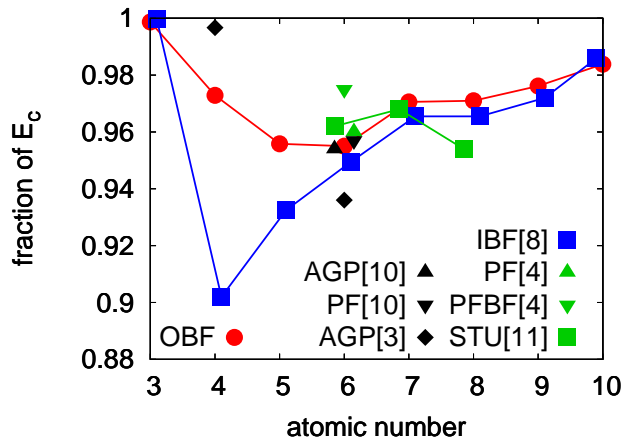}
\caption{Fraction of correlation energy recovered in fixed--node DMC
for the first row atoms of orbital backflow wave functions (OBF) compared to
previous results of different kinds of single--determinant wave functions.
Notations are the same as those of
Fig. \ref{fig_vmc}. 
}
\label{fig_dmc}
\end{figure}

Energies of the first row atoms have been calculated 
by several authors
using a variety of different trial wave functions beyond the simple JS form
providing useful comparisons.
In Figs. \ref{fig_vmc} and \ref{fig_dmc} our OBF data from Table 
\ref{tab_table}, indicated by full red circles,
are compared with selected results
from the literature, as indicated by the labels in the body of the
figures with the reference in brackets (unpublished
calculations \cite{sorella_unpublished} using AGP and PF, and
an earlier AGP result from Ref. \onlinecite{sorella}; the IBF energies
from Ref. \onlinecite{seth} for VMC in Fig. \ref{fig_vmc} and 
from Ref. \onlinecite{brown} for DMC in Fig. \ref{fig_dmc};
a PF calculation \cite{schmidt_prb} and its version (PFBF) with 
IBF included, and a general PF form dubbed STU 
\cite{schmidt_prl} which encompasses  
both singlet and triplet pairing, as well as unpaired orbitals).
For AGP and IBF, subsequent calculations with the same kind of
wave function found lower energies on account of more aggressive 
optimization and/or use of extended basis sets. 
We also 
show by empty symbols in Fig. \ref{fig_vmc} the VMC JS result from 
the respective sources.
Orbital dependent Jastrow correlations 
applied to the oxygen atom \cite{caffarel2} have not led to 
significant improvement compared to the common JS trial wave function within VMC and DMC.

The most pertinent and systematic comparison is possible between IBF and OBF.
Both are ways to include backflow effects in inhomogeneous systems,
where standard backflow of the form of Eq. (\ref{eq_stdbf}) with the
simple coordinate transformation of Eq. (\ref{eq_eta}) does not
significantly lower the energies.
Within IBF, nuclear coordinates are included inside
the standard BF transformation through {\em atom--specific} 
electron--nucleus
\cite{mh_bf,lopezrios} and electron--electron--nucleus \cite{brown} terms.
Instead, OBF only uses the basic BF transformation with homogeneous
electron-electron term, but introduces an {\em orbital--specific}
dependence through the modified orbitals of Eq. (\ref{eq_obf}). 
We mention that a yet different BF wave function, featuring iterative 
coordinate renormalization \cite{bfiter1}, gives excellent results for both
homogeneous and inhomogeneous strongly
correlated systems \cite{bfiter2}. 
However, it becomes less beneficial as correlations
weaken, providing only marginal improvements for the first row atoms.

Whereas OBF obtains slightly less correlation energy than IBF\cite{seth} within VMC
(see Fig.~\ref{fig_vmc}), a small gain relative to IBF is obtained by OBF at the level of fixed-node DMC,
except for lithium and neon
where they are very close.
We attribute the qualitatively different behavior of VMC and DMC 
to a better parametrization of the symmetric Jastrow factor of the IBF in Ref.\onlinecite{seth}
compared to the Jastrow factor of the present work, clearly visible in
the difference in the respective bare Slater-Jastrow (JS) data
(see Fig. \ref{fig_vmc}). 
We further note that 
the DMC results for single--determinant IBF 
are only provided by Ref. \onlinecite{brown}.
In Ref.~\onlinecite{seth}, the
IBF wave function was optimized much better, lowering the VMC energies 
particularly in the case of beryllium and boron, but DMC values have not been provided.
It is natural to expect that the better
optimized IBF wave function will also lower the corresponding DMC values, thus reducing the
rather large difference between IBF and OBF of those 
two atoms shown in  Fig.~\ref{fig_dmc}. 
For the other atoms, however, the variational
quality of single--determinant IBF wave functions given in Refs. 
\onlinecite{brown} and \onlinecite{seth} is very similar, and the DMC results  
of Ref.~\onlinecite{brown} shown in Fig. \ref{fig_dmc} should be representative
of a well--optimized IBF.
Overall, it seems fair to conclude that the OBF nodes tend to provide  a slightly better
description than those of IBF.

Pairing wave functions (AGP, PF or STU) 
overcome some topological inadequacies \cite{mitas_nodes}
that Hartree-Fock nodes share with single--determinant wave 
functions, including OBF and IBF. The excellent result
--particularly in DMC-- obtained for beryllium with the AGP wave 
function is due to its multi-determinant character\cite{sorella} with respect to 
the nearly--degenerate $p$ orbitals. However, for the heavier atoms,
single--determinant wave functions with OBF or IBF provide essentially as good energies as
pairing wave functions. The further, non--negligible gain obtained by
inclusion of IBF in a PF wave function \cite{schmidt_prb} 
(see the PFBF points in Figs. \ref{fig_vmc} and \ref{fig_dmc})
suggests that pairing and backflow correlations improve complementary
aspects of the wave function, at least to some
extent.

So far, the discussion and our comparison in Figs. \ref{fig_vmc} and
\ref{fig_dmc} has been restricted exclusively to single--determinant trial wave functions.
For small atoms,
nearly exact energies can be retrieved in QMC multi-determinant expansions with a modest 
number of terms
\cite{seth,morales,toulouse}. However, the 
same accuracy cannot be maintained for heavier atoms or molecules
with an affordable number of determinants, whereas backflow wave functions
should improve the accuracy fairly independent of the number of electrons
with only polynomial increasing computational cost.

We briefly mention that a simple linear extrapolation of the JS and OBF energies 
against the corresponding variancies of the energy, as succesfully used in strongly correlated
homogeneous quantum fluids \cite{bfiter1,bfiter2}, does not provide any systematic improvement
for the first row atoms' energies. The discrete nature of the density of states for the electrons
in the nuclear potential seems to considerably shrink down the region of validity for such extrapolations.

\section{Discussion}
Orbital--dependent backflow wave functions provide
a simple and efficient way of introducing
and tuning a physically appealing orbital dependence in many--body correlations.
We have shown that the resulting gain in energy for first row atoms is
competitive with inhomogeneneous backflow wave functions \cite{seth},
the best currently available single--determinant trial wave function for electronic
structure of atoms and molecules.
For more complex systems, the orbital dependence of OBF presents an appealing 
alternative to the atom--specific IBF transformation, and may be better suited
to study orbital--selective phenomena in strongly correlated systems
\cite{selectivemott}.

Variational flexibility of OBF is added by two main ingredients,
a coordinate renormalization, Eq. (\ref{eq_eta}), and an
orbital modification, Eq. (\ref{eq_obf}). The latter could be
used without the former, for instance replacing the backflow coordinate by
the fluctuation of a local dipole
or by the wave vector of a density fluctuation in an extended system
in the scalar product with the gradient.
In such a version, OBF would correspond to an earlier 
representation \cite{tocchio} of backflow correlations used in lattice models \cite{luo}.

Generically, OBF provides a modification of orbitals enlarging the functional
flexibility of trial wave functions suited for standard real space QMC methods.
It can directly be combined with IBF including further dependency on the
electron--nucleus distances or electron--electron--nucleus  in the backflow coordinates,
and extended to iterated backflow wave functions, as well as in the use of Pfaffian
and multi--determinant trial wave function. Efficient optimization among all possible 
combinations may request improved optimization strategies \cite{kochkov}.

Despite obvious limitations, the accuracy reached by real--space QMC methods
should be sufficient to tackle and provide new insights to the role of
correlation in electronic structure. Flexible trial wave functions capturing different aspects
of correlation
put up a frame to estimate and reduce the bias of the underlying trial wave function.
Still, reliabe estimates and/or control of the
the systematic error of 
fixed--node QMC involving large number of electrons remains challenging.

\section*{Acknowledgment}
We thank the Fondation NanoSciences (Grenoble) and the CNRS, INP for support.


\begin{thebibliography}{99}
\bibitem{reviews} J. Koloren\v{c}. and L. Mitas, Rep. Prog. Phys. {\bf 74}, 026502 (2011).
\bibitem{filippi} C. Filippi and C. J. Umrigar, J. Chem. Phys. 105, 213 (1996).
\bibitem{sorella} M. Casula, C. Attaccalite, and S. Sorella, 
J. Phys. Chem. 121, 7110 (2004).
\bibitem{schmidt_prb} M. Bajdich, L. Mitas, L. K. Wagner and K. E. Schmidt, 
Phys. Rev. B {\bf 77}, 115112 (2008).
\bibitem{slkc} K. E. Schmidt, M. A. Lee, M. H. Kalos, and G. V. Chester, 
Phys. Rev. Lett. 47, 807 (1981).
\bibitem{kwon} Y. Kwon, D. M. Ceperley and R. M. Martin, 
Phys. Rev. B {\bf 48}, 12037 (1993).
\bibitem{mh_bf} M. Holzmann, D. M. Ceperley, C. Pierleoni and K. Esler, 
Phys. Rev. E {\bf 68}, 046707 (2003).
\bibitem{brown} M. D. Brown, J. R. Trail, P. L\'opez R\'{\i}os, and R. J. Needs, 
J. Chem. Phys. 126, 224110 (2007).
\bibitem{seth} P. Seth, P. L\'opez R\'{\i}os, and R. J. Needs
J. Chem. Phys. 134, 084105 (2011).
\bibitem{sorella_unpublished} S. Sorella, private communication.
\bibitem{schmidt_prl} M. Bajdich, L. Mitas, G. Drobn\'y, L. K. Wagner, 
and K. E. Schmidt, Phys. Rev. Lett. 96, 130201 (2006).
\bibitem{kalos} K. S. Liu, M. H. Kalos, and G. V. Chester, 
Phys. Rev. A {\bf 10}, 303 (1974).
\bibitem{caffarel} M. Caffarel and P. Claverie, 
J. Chem. Phys. {\bf 88}, 1088 (1988).
\bibitem{natoli} V. Natoli and D. M. Ceperley, 
J. Comput. Phys. {\bf 117}, 171 (1995).
\bibitem{roetti} E. Clementi and C. Roetti, 
At. Data Nucl. Data Tables {\bf 14}, 177 (1974).
\bibitem{rocca} S. Sorella, M. Casula, and D. Rocca, 
J. Chem. Phys. 127, 014105 (2007).
\bibitem{umrigar} C. J. Umrigar, J. Toulouse, C. Filippi, S. Sorella, 
and R. G. Hennig, Phys. Rev. Lett. {\bf 98}, 110201 (2007).
\bibitem{caffarel2} T. Bouab\c{c}a, B. Bra{\"i}da, and M. Caffarel, J. Chem. Phys. {\bf 133}, 044111 (2010).
\bibitem{lopezrios} P. L\'opez R\'{\i}os, A. Ma, N. D. Drummond, M. D. Towler, 
and R. J. Needs, Phys. Rev. E {\bf 74}, 066701 (2006).
\bibitem{bfiter1} M. Taddei, M. Ruggeri, S. Moroni, and M. Holzmann,
Phys. Rev. B {\bf 91}, 115106 (2015).
\bibitem{bfiter2} M. Ruggeri, S. Moroni, and M. Holzmann,
Phys. Rev. Lett. {\bf 120}, 205302 (2018).
\bibitem{mitas_nodes} K. M. Rasch and L. Mitas,
Chem. Phys. Lett. {\bf 528}, 59 (2012).
\bibitem{morales} M. A. Morales, J. McMinis, B. K. Clark, J. Kim, and G. E. Scuseria,
J. Chem. Theory Comput. {\bf 8} (7), 2181 (2012). 
\bibitem{toulouse} J. Toulouse and C. J. Umrigar, J. Chem. Phys. {\bf 128}, 174101 (2008).
\bibitem{selectivemott} See, e.g., V. I. Anisimov, 
I. A. Nekrasov, D. E. Kondakov, T. M. Rice and M. Sigrist,
Eur. Phys. J. B 25, 191-201 (2002).
\bibitem{tocchio} L. F. Tocchio, F. Becca, A. Parola, and S. Sorella,
Phys. Rev. B {\bf 78}, 041101(R) (2008).
\bibitem{luo} Di Luo and B K. Clark,  arXiv:1807.10770.
\bibitem{kochkov} D. Kochkov and B. K. Clark, arXiv:1811.12423.
\end{thebibliography}
\end{document}